\documentclass[final,5p,times,twocolumn]{elsarticle}

\usepackage{amssymb}
\journal{   }

\begin{document}

\begin{frontmatter}

%% Title, authors and addresses

%% use the tnoteref command within \title for footnotes;
%% use the tnotetext command for theassociated footnote;
%% use the fnref command within \author or \address for footnotes;
%% use the fntext command for theassociated footnote;
%% use the corref command within \author for corresponding author footnotes;
%% use the cortext command for theassociated footnote;
%% use the ead command for the email address,
%% and the form \ead[url] for the home page:
%% \title{Title\tnoteref{label1}}
%% \tnotetext[label1]{}
%% \author{Name\corref{cor1}\fnref{label2}}
%% \ead{email address}
%% \ead[url]{home page}
%% \fntext[label2]{}
%% \cortext[cor1]{}
%% \address{Address\fnref{label3}}
%% \fntext[label3]{}

\title{Coherent photoproduction of low-$p_{T}$ charmonium in peripheral heavy ion collisions within the color dipole model}

%% use optional labels to link authors explicitly to addresses:
%% \author[]{}
%% \address[label1]{}
%% \address[label2]{}

\author[HRBEU]{Gongming Yu}
%\author{Gong-Ming Yu\corref{mycorrespondingauthor}}
%\cortext[mycorrespondingauthor]{Corresponding author}
\ead{ygmanan@163.com}
\address[HRBEU]{Fundamental Science on Nuclear Safety and Simulation Technology Laboratory, Harbin Engineering University, Harbin 150000, China}

\author[GZUFE]{Yanbing Cai}
\ead{myparticle@163.com}
\address[GZUFE]{Guizhou Key Laboratory in Physics and Related Areas, and Guizhou Key Laboratory of Big Data Statistic Analysis, Guizhou University of Finance and Economics, Guiyang 550025, China}

\author[CYNU]{Yongping Fu}
%\author{Gong-Ming Yu\corref{mycorrespondingauthor}}
%\cortext[mycorrespondingauthor]{Corresponding author}
\ead{ynufyp@sina.cn}
\address[CYNU]{department of physics, West Yunnan University, Lincang 677000, China}

\author[ZTU]{Haitao Yang}
%\author{Gong-Ming Yu\corref{mycorrespondingauthor}}
%\cortext[mycorrespondingauthor]{Corresponding author}
\ead{yanghaitao205@163.com}
\address[ZTU]{Department of Physics, Zhaotong University, Zhaotong 657000, China}

\author[YXNU]{Quangui Gao}
%\author{Gong-Ming Yu\corref{mycorrespondingauthor}}
%\cortext[mycorrespondingauthor]{Corresponding author}
\ead{qggao@yxnu.edu.cn}
\address[YXNU]{Department of Physics, Yuxi Normal University, Yuxi 653100, China}

\author[IMP]{Qiang Hu}
\ead{qianghu@impcas.ac.cn}
\address[IMP]{Institute of Modern Physics, Chinese Academy of Sciences, Lanzhou 730000, China}

\author[HRBEU]{Liyuan Hu}
%\author{Gong-Ming Yu\corref{mycorrespondingauthor}}
%\cortext[mycorrespondingauthor]{Corresponding author}
%\ead{huliyuan@hrbeu.edu.cn}
%\address[HRBEU]{Fundamental Science on Nuclear Safety and Simulation Technology Laboratory, Harbin Engineering University, Harbin 150000, China}

\author[HRBEU]{Wei Li}
%\author{Gong-Ming Yu\corref{mycorrespondingauthor}}
%\cortext[mycorrespondingauthor]{Corresponding author}
\ead{liw-official@163.com}
%\address[HRBEU]{Fundamental Science on Nuclear Safety and Simulation Technology Laboratory, Harbin Engineering University, Harbin 150000, China}

\author[HRBEU]{Yushou Song}
%\author{Gong-Ming Yu\corref{mycorrespondingauthor}}
%\cortext[mycorrespondingauthor]{Corresponding author}
\ead{yushousong@hrbeu.edu.cn}
%\address[HRBEU]{Fundamental Science on Nuclear Safety and Simulation Technology Laboratory, Harbin Engineering University, Harbin 150000, China}

%\author{Larry D. McLerran}
%\author{Larry D. McLerran\corref{mycorrespondingauthor}}
%\cortext[mycorrespondingauthor]{Corresponding author.}
%\ead{mclerran@me.com}

%\address{Institute for Nuclear Theory, University of Washington, Box 351550, Seattle, WA 98195, USA}

%%\cortext[mycorrespondingauthor]{Corresponding author}
%%\ead{support@elsevier.com}

\begin{abstract}
We calculate the centrality dependence for coherent photoproduction of very low-$p_{T}$ $J/\psi$ at Relativistic Heavy Ion Collider (RHIC) and Large Hadron Collider (LHC) energies within the impact parameter dependent saturated color dipole model. By using the large equivalent photon fluxes, we present the differential cross section of very low-$p_{T}$ $J/\psi$ produced by coherent photonuclear in peripheral heavy-ion collisions. The numerical results demonstrate that our calculation are agree with $J/\psi$ data in peripheral heavy ion collisions at Relativistic Heavy Ion Collider (RHIC) energies.
\end{abstract}

%\begin{keyword}

%Glasma; Quark-Gluon plasma; Hadronic gas; Electromagnetic production

%PACS number(s): 25.75.-q, 21.65.Qr, 12.38.Mh, 13.85.Qk

%% keywords here, in the form: keyword \sep keyword
%% PACS codes here, in the form: \PACS code \sep code
%% MSC codes here, in the form: \MSC code \sep code
%% or \MSC[2008] code \sep code (2000 is the default)

%\end{keyword}

\end{frontmatter}

%% \linenumbers

%% main text

\section{Introduction}

The Relativistic Heavy Ion Collider (RHIC) and Large Hadron Collider (LHC) are the most powerful collider for the central and non-central heavy ion collisions, offering a unique opportunity to study fundamental aspects of Quantum Electrodynamics (QED) and Quantum Chromodynamics (QCD). The central collisions of heavy ions provide a unique tool to create and study the strongly interacting  matter, known as quark-gluon plasma (QGP) at high energy density and temperature \cite{1,2,3,4,5,6,7,8,9}. The $J/\psi$ suppression in heavy-ion collisions has been proposed as a signature of QGP formation \cite{7}. The peripheral collisions of heavy ions at RHIC and LHC give an opportunity to explore high-energy nuclear physics with beams of quasi-real photons \cite{10,11,12,13,14}. The $J/\psi$ can also be produced via the strong electromagnetic fields generated by heavy ions, e.g. photon-nucleus coherent or incoherent interactions, in peripheral heavy-ion collisions \cite{15,16,17,18,19,20,21}. Recently, a significant excess of $J/\psi$ yield at very low transverse momenta has been observed by the ALICE collaboration in peripheral hadronic Pb-Pb collisions with $\sqrt{s_{NN}}=2.76TeV$ at forward-rapidity \cite{22}, and by the STAR collaboration in hadronic Au-Au collisions with $\sqrt{s_{NN}}=200GeV$ and U-U collisions with $\sqrt{s_{NN}}=193GeV$ at mid-rapidity \cite{23,24}, that cannot be explained within the hadronic $J/\psi$ production modified by the cold and hot medium effects. It indicates that the significant excess maybe originated from the coherent photoproduction in hadronic heavy-ion collisions ($b<2R$). In this process, the strong electromagnetic fields generated by the colliding ions can be represented by a spectrum of equivalent photons, that can be used to study coherent photonuclear interactions. The quasireal photons coherently interact with the gluon field of the other nucleus to produce a $J/\psi$ with low transverse momentum.

In the present work, we investigate the coherent photoproduction of very low-$p_{T}$ $J/\psi$ within the impact parameter dependent saturated dipole model, that can be described by dipole-nucleus scattering amplitude. The dipole model became an important tool in investigations of deep-inelastic scattering due to the simple ansatz for the dipole cross section integrated over the impact parameter ($\mathbf{b}$), that  was able to describe simultaneously the total inclusive and diffractive cross sections. In the color dipole model, photon-nucleus scattering process is described as the virtual photon fluctuating into a quark-antiquark color dipole, that then scatters off the nucleus, via Pomeron exchange, the perturbative-QCD equivalent of which is the exchange of gluon ladder. Coherently produced $J/\psi$ in peripheral hadronic collisions are expected to probe the nuclear gluon distribution at low Bjorken $x$ which is still considerable uncertainty.

The paper is organized as follows. In Sec. 2, we present the coherent photoproduction of very low-$p_{T}$ $J/\psi$ in peripheral heavy ion collisions within the impact parameter dependent saturated dipole model. The numerical results for low-$p_{T}$ $J/\psi$ in Au-Au and U-U collisions at RHIC energies, and Pb-Pb collisions at LHC energies are plotted in Sec. 3. Finally, the conclusion is given in Sec. 4.

%The paper is organized as follows. In Section 2, we present the
%production of low-mass dileptons and low-$p_{T}$ photons from the
%Glasma, quark-gluon plasma(QGP) and hadronic gas(HG) at Relativistic
%Heavy Ion Collider(RHIC) and Large Hadron Collider(LHC) energies,
%based on the relativistic kinetic theory. The numerical results for
%Au-Au collisions at Relativistic Heavy Ion Collider(RHIC) and Pb-Pb
%collisions at Large Hadron Collider(LHC) energies are plotted in
%Section 3. Finally, the conclusion is given in Section 4.

\section{General formalism}

In peripheral nucleus-nucleus collisions, the strong interactions are heavily suppressed, and the electromagnetic interaction is expected to dominate. The differential cross-section for coherent photoproduction of very low-$p_{T}$ charmonium in peripheral heavy ion collisions within the impact parameter dependent saturated dipole model can be written as
\begin{eqnarray}\label{y1}
\!\!\!\!\!\!\!\!\!\!\!\!d\sigma\!\!\!\!&=&\!\!\!\!dN_{\gamma}(\mathbf{r},\omega)d\hat{\sigma}_{\gamma^{\ast}A\rightarrow J/\psi A}(|\mathbf{r}-\mathbf{b}|,x_{p},Q^{2},\Delta)\nonumber\\[1mm]
\!\!\!\!\!\!\!\!\!\!\!\!\!\!\!\!&=&\!\!\!\!d^{2}rd\omega\frac{dN_{\gamma}(\mathbf{r},\omega)}{d^{2}rd\omega}d\hat{t}d^{2}b\frac{d\hat{\sigma}_{\gamma^{\ast}A\rightarrow J/\psi A}}{d\hat{t}d^{2}b}(|\mathbf{r}-\mathbf{b}|,x_{p},Q^{2},\Delta),
\end{eqnarray}
where $\mathbf{r}$ and $\mathbf{b}$ are the impact parameter, and $\hat{t}=-\Delta^{2}$ is the transfer momentum. The energy for the photon is $\omega=\frac{M_{V}}{2}\exp(y)$, here $M_{V}$ and $y$ are the vector meson mass and rapidity, respectively. In the center-of-mass frame, the transformation $d\hat{t}\sim dp_{T}^{2}$ and $d\omega=\omega dy$ can be performed. Therefore the differential cross section for the nucleus-nucleus collisions can be written in the terms of charmonium  transverse momentum as the following
\begin{eqnarray}\label{y2}
\!\!\!\!\!\!\!\!\!\!\!\!\frac{d\sigma}{dp_{T}^{2}dy}=\!\!\int d^{2}rd^{2}b\frac{dN_{\gamma}(\mathbf{r},\omega)}{d^{2}rd\omega}\frac{d\hat{\sigma}_{\gamma^{\ast}A\rightarrow J/\psi A}}{d\hat{t}d^{2}b}(|\mathbf{r}-\mathbf{b}|,x_{p},Q^{2},\Delta),\!\!\!\!
\end{eqnarray}
here $x_{p}=\frac{M_{V}}{2}\exp(y)/\sqrt{s_{NN}}$ is the momentum fraction of the gluon probed by the photon.

The equivalent photon spectrum for nucleus can be obtained from the semiclassical description of high-energy electromagnetic collisions. A relativistic nucleus with $Z$ times the electric charge moving with a relativistic factor $\gamma_{L}\gg1$ with respect to develop an equally strong magnetic field component hence it resembles a beam of photons, where the number of photons can be expressed as \cite{25,26,27}
\begin{eqnarray}\label{y3}
\frac{dN_{\gamma}(\mathbf{r},\omega)}{d^{2}rd\omega}=\frac{Z^{2}\alpha\eta^{2}}{\pi^{2}r^{2}\omega}\left[K_{1}(\eta)+\frac{1}{\gamma_{L}^{2}}K_{0}^{2}(\eta)\right],
\end{eqnarray}
where $\eta=\omega r/\gamma_{L}$, $\omega$ is the photon momentum, $K_{0}(x)$ and $K_{1}(x)$ are the Bessel function, and $\alpha$ is the electromagnetic coupling constant.

The differential cross section for the quasielastic coherent vector meson photoproduction in nucleus-nucleus collisions can be written as \cite{28,29}
\begin{eqnarray}\label{y4}
\!\!\!\!\!\!\!\!\!\!\!\!\frac{d\hat{\sigma}_{\!\!\gamma^{\ast}A\rightarrow\! VA}}{d\hat{t}d^{2}b}(|\mathbf{r}\!-\!\mathbf{b}|,\!x_{p},\!Q^{2}\!,\!\Delta)\!\!\!\!\!&=&\!\!\!\!\!\!\frac{R_{g}^{2}(1\!+\!\beta^{2})}{16\pi}\left|A_{T,L}^{\gamma^{\ast}A\rightarrow VA}\!(|\mathbf{r}\!-\!\mathbf{b}|,\!x_{p},\!Q^{2}\!,\!\Delta)\right|^{2}\!\!\!,\nonumber\\[1mm]
&&\!\!\!\!\!\!\!
\end{eqnarray}
with
\begin{eqnarray}\label{y5}
\!\!\!\!\!\!\!\!R_{g}=\frac{2^{2\delta+3}}{\sqrt{\pi}}\frac{\Gamma(\delta+5/2)}{\Gamma(\delta+4)},\beta=\tan(\frac{\pi\delta}{2}),\delta=\frac{\partial\ln A_{T,L}^{\gamma^{\ast}A\rightarrow VA}}{\partial\ln(1/x_{p})},
\end{eqnarray}
the elementary elastic amplitude $A_{T,L}^{\gamma^{\ast}A\rightarrow VA}(|\mathbf{r}-\mathbf{b}|,x_{p},Q^{2},\Delta)$ which is defined such that the elastic differential cross section for the quark-antiquark color dipole scattering on the nucleus is given by
\begin{eqnarray}\label{y6}
\!\!\!\!\!\!\!\!\!\!\!\!A_{T,L}^{\gamma^{\ast}A\rightarrow VA}(|\mathbf{r}-\mathbf{b}|,x_{p},Q^{2},\Delta)\!\!\!\!&=&\!\!\!\!\!\!\int\frac{dz}{4\pi}d^{2}r_{d}e^{-i[|\mathbf{r}-\mathbf{b}|-(1-z)\mathbf{r}_{d}]*\Delta}\nonumber\\[1mm]
&&\!\!\!\!\!\!\!\!\!\!\!\times\left(\Psi_{V}^{\ast}\Psi\right)_{T,L}\frac{d\hat{\sigma}_{q\bar{q}}}{d^{2}b}(|\mathbf{r}-\mathbf{b}|,\mathbf{r}_{d},x_{p}),
%&&\!\!\!\!\times 2\left\{1-\exp\left[2\pi B_{p}AT_{A}(b)N(r_{d},x_{p})\right]\right\},
\end{eqnarray}
here $A$ is the nucleon number, the overlaps between the photon and the vector meson wave functions can be written as \cite{28}
\begin{eqnarray}\label{y7}
(\Psi_{V}^{\ast}\Psi)_{T}\!\!\!\!&=&\!\!\!\!\hat{e}_{f}e\frac{N_{c}}{\pi z(1-z)}\big\{m_{f}^{2}K_{0}(\epsilon r_{d})\phi_{T}(r_{d},z)\nonumber\\[1mm]
&&\!\!\!\!-[z^{2}+(1-z)^{2}]\epsilon K_{1}(\epsilon r_{d})\partial_{r_{d}}\phi_{T}(r_{d},z)\big\},
\end{eqnarray}
\begin{eqnarray}\label{y8}
(\Psi_{V}^{\ast}\Psi)_{L}\!\!\!\!&=&\!\!\!\!\hat{e}_{f}e\frac{N_{c}}{\pi}2Qz(1-z)K_{0}(\epsilon r_{d})\big\{m_{v}^{2}\phi_{T}(r_{d},z)\nonumber\\[1mm]
&&\!\!\!\!+\delta\frac{m_{f}^{2}-\nabla_{r_{d}}^{2}}{M_{v}z(1-z)}\phi_{L}(r_{d},z)\big\},
\end{eqnarray}
where $N_{c}=3$, $\hat{e}_{f}=2/3$, $e=\sqrt{4\pi\alpha}$, $\epsilon^{2}=z(1-z)Q^{2}+m_{f}^{2}$, $m_{f}$ is the quark mass, and $M_{v}$ is the mass of vector meson. The scalar functions $\phi_{T,L}(r_{d},z)$ of vector meson, that is parameterized in the boosted Gaussian expressions, have the following general form \cite{30,31,32,33,34}
\begin{eqnarray}\label{y9}
\!\!\!\!\!\!\!\!\!\!\!\!\phi_{T,L}^{J/\psi}(r_{d},z)\!=\!\!\mathcal{N}_{T,L}z(1\!-\!z)\exp\bigg(\!\!-\!\frac{m_{f}^{2}\mathcal{R}^{2}}{8z(1\!-\!z)}\!-\!\frac{2z(1\!-\!z)r_{d}^{2}}{\mathcal{R}^{2}}\!+\!\frac{m_{f}^{2}\mathcal{R}^{2}}{2}\!\bigg),\!\!\!\!\!\!
\end{eqnarray}
here the parameters ($\mathcal{N}_{T,L}$ and $\mathcal{R}^{2}$) of the boosted Gaussian scalar functions for $J/\psi$ can be found in Ref. \cite{28}.

For a large and smooth nucleus, the dipole-nucleus cross section $\frac{d\hat{\sigma}_{q\bar{q}}}{d^{2}b}(|\mathbf{r}-\mathbf{b}|,\mathbf{r}_{d},x_{p})$ in Glauber-Gribov approach is given by \cite{35,36,37,38,39,40}
\begin{eqnarray}\label{y10}
&&\!\!\!\!\!\!\!\!\!\!\!\!\!\!\!\!\!\!\!\!\!\!\!\!\frac{d\hat{\sigma}_{q\bar{q}}}{d^{2}b}(|\mathbf{r}\!-\!\mathbf{b}|,\mathbf{r}_{d},x_{p})\!=\!2\left\{1\!-\!\exp\left[2\pi B_{p}AT_{A}(|\mathbf{r}\!-\!\mathbf{b}|)\hat{\sigma}_{q\bar{q}}(r_{d},x_{p})\right]\right\},\nonumber\\[1mm]
&&
\end{eqnarray}
here the diffractive slope parameters ($B_{p}$) and the nuclear profile function $T_{A}(|\mathbf{r}-\mathbf{b}|)$ can be found in Ref. \cite{41,42,43}. There are many dipole cross-section parametrizations available in the literature, and we have taken for this study three representative models. In the Golec-Biernat and W$\ddot{u}$sthoff (GBW) model, the dipole cross-section can be written as \cite{44,45}
\begin{eqnarray}\label{y11}
\hat{\sigma}_{q\bar{q}}^{GBW}(r_{d},x_{p})=\sigma_{0}\left(1-e^{-r_{d}^{2}Q_{s}^{2}(x_{p})/4}\right),
\end{eqnarray}
where $\sigma_{0}$ is a constant and $Q_{s}^{2}(x_{p})$ denotes the $x_{p}$ dependent saturation scale \cite{28}.  Despite the appealing simplicity and success of the GBW model, it suffers from clear shortcomings. In particular, it does not include scaling violations, that isoes not match with QCD Dokshitzer-Gribov-Lipatov-Altarelli-Parisi (DGLAP) evolution at
large $Q^{2}$. Therefore, Bartels, Golec-Biernat, and Kowalski (BGBK) proposed a new the dipole cross-section by replacing $Q^{2}$ by a gluon density with explicit DGLAP evolution \cite{46}
\begin{eqnarray}\label{y12}
\!\!\!\!\!\!\!\!\!\!\!\!\hat{\sigma}_{q\bar{q}}^{BGBK}\!(r_{d},x_{p})\!=\!\sigma_{0}\!\left\{1\!-\!\exp\left[-\pi^{2}r_{d}^{2}\alpha_{s}(\mu^{2})xg(x_{p},\mu^{2})/(3\sigma_{0})\right]\right\},\!
\end{eqnarray}
here $xg(x_{p},\mu^{2})$ is gluon density at gluon density scale $\mu^{2}=C/r_{d}^{2}+\mu_{0}^{2}$ \cite{28}. But the DGLAP evolution may not be appropriate when the $x$ approach to the saturation region. Therefore, Iancu, Itakura, and Munier proposed a new saturation model, the color glass condensate (CGC) model \cite{47}, that can be written as
\begin{eqnarray}\label{y13}
\!\!\!\!\!\!\!\!\!\!\!\!\hat{\sigma}_{q\bar{q}}^{CGC}(r_{d},x_{p})=\sigma_{0}
\left\{
             \begin{array}{lr}
             \!\!\!\mathcal{N}_{0}\left(\frac{r_{d}}{Q_{s}}\right)^{2\left[\gamma_{s}+(1/\kappa\lambda Y)\ln(2/r_{d}Q_{s})\right]}, & \!\!\!\!\!r_{d}Q_{s}\leq2, \\
             \!\!\!1-e^{-A\ln^{2}\left(Br_{d}Q_{s}\right)}, & \!\!\!\!\!r_{d}Q_{s}>2,
             \end{array}
\right.\!\!\!\!
\end{eqnarray}
where $Y=\ln(1/x_{p})$, $\gamma_{s}=0.63$, and $\kappa=9.9$. The coefficients $A$ and $B$ are determined uniquely by matching of the two parts of dipole amplitude and their logarithmic derivatives at $r_{d}Q_{s}=2$, and the free parameters $\sigma_{0}$ and $\mathcal{N}_{0}$ were determined by a fit to Hadron Electron Ring Accelerator (HERA) $F_{2}$ data \cite{28}.

The saturation scale $Q_{s}$ which depends on the impact parameter is given by
\begin{eqnarray}\label{y14}
Q_{s}=Q_{s}(x,b)=\left(\frac{x_{0}}{x}\right)^{\lambda/2}\left[\exp\left(-\frac{b^{2}}{B_{CGC}}\right)\right]1/2\gamma_{s},
\end{eqnarray}
where the value of $B_{CGC}=5.5GeV^{-2}$ is derived by the $t$ distribution of the exclusive diffractive processes at
Hadron Electron Ring Accelerator (HERA) \cite{29}.

\section{Numerical results}

In order to present our results in a way that can be compared with STAR data from RHIC, we will study the invariant yield of very low-$p_{T}$ $J/\psi$ production as the following
\begin{eqnarray}\label{y15}
B_{e^{+}e^{-}}\frac{dN}{dp_{T}^{2}dy}=\frac{B_{e^{+}e^{-}}}{\sigma_{total}}\frac{d\sigma}{dp_{T}^{2}dy},
\end{eqnarray}
where $B_{e^{+}e^{-}}=5.97\pm0.03\%$ is the branching ratio for $J/\psi$ decay into electron-position pair ($J/\psi\rightarrow e^{+}e^{-}$) \cite{48}, $\sigma_{total}\sim4\pi R_{T}^{2}$ is the total cross section for nucleus-nucleus collisions \cite{49}, $R_{T}=1.2A^{1/3}fm$ is the transverse radius of the nucleus, and $A$ is the nucleon number of the nucleus.

\begin{figure}
\includegraphics[width=9cm,height=7cm]{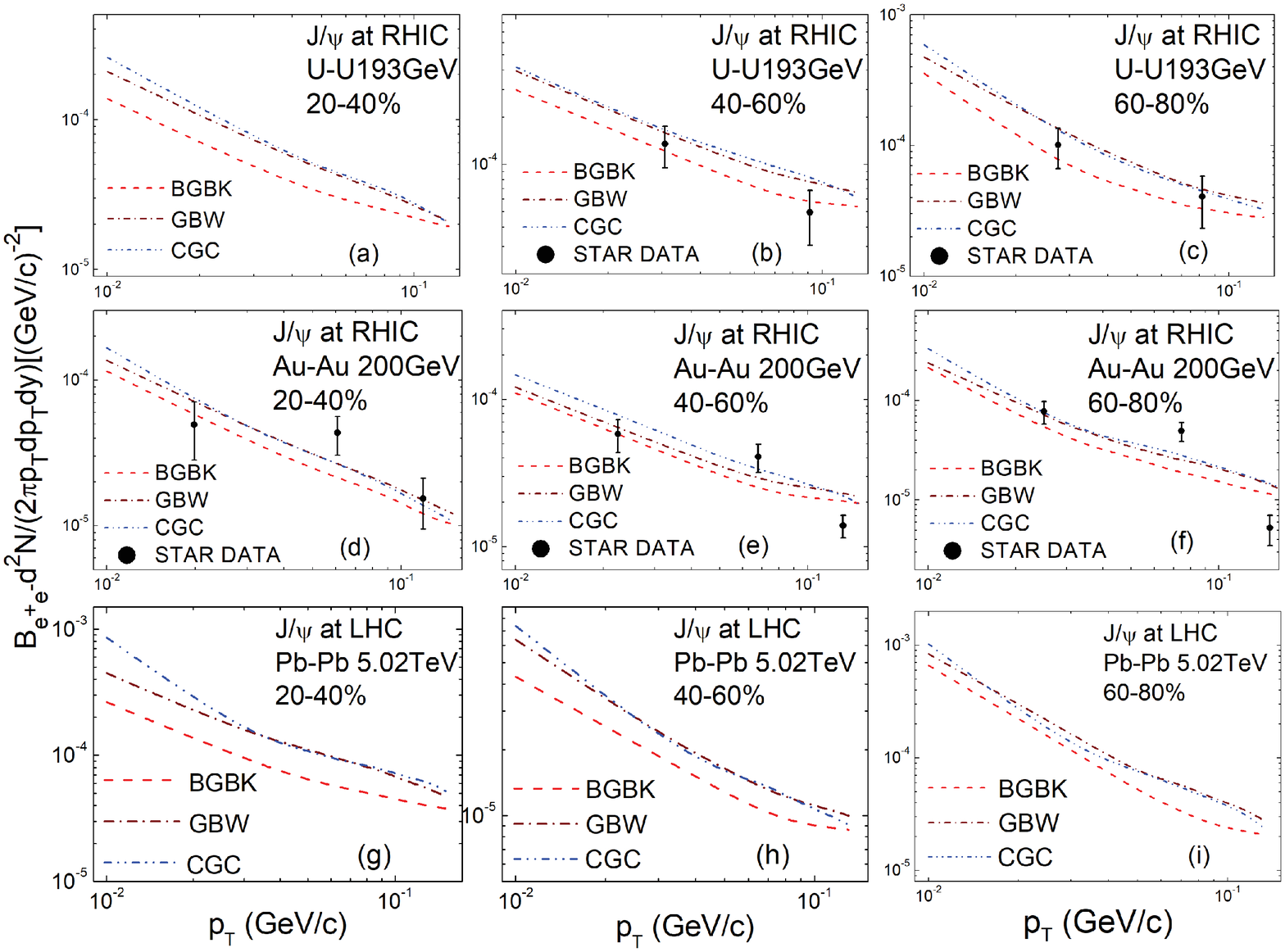}
\caption{The very low-$p_{T}$ $J/\psi$ yield in relativistic heavy ion collisions. The dash line (red) is for the GBW model, the dash-dot line (wine) is for the BGBK model, the dash-dot-dot line (blue) is for the CGC model. The data points are from the
STAR Collaboration \cite{24}.}\label{ygm1}
\end{figure}

In Fig.\,\ref{ygm1}, we plot the spectra of low-$p_{T}$ $J/\psi$ produced by coherent photoproduction processes in Au-Au collisions with $\sqrt{s_{NN}}=200\mathrm{GeV}$ and U-U collisions with $\sqrt{s_{NN}}=193\mathrm{GeV}$ at RHIC, as well as Pb-Pb collisions with $\sqrt{s_{NN}}=5.02\mathrm{TeV}$ at LHC. Compared with STAR Collaboration $J/\psi$ meson data \cite{24}, we can see that our results are agree with the coherently produced $J/\psi$ at very low transverse momentum.

%%%%%%%%%%%%%%%%%%%%%%%%%%%%%%%%%%%%%%%%%%%%%%%%%%%%%%%%%%%%%%%%%%%%%%%%%%%%%%%%%%%%%%%%%%%%%%%%%%%%%%%%%%%%%%
%%%%%%%%%%%%%%%%%%%%%%%%%%%%%%%%%%%%%%%%%%%%%%%%%%%%%%%%%%%%%%%%%%%%%%%%%%%%%%%%%%%%%%%%%%%%%%%%%%%%%%%%%%%%%%

\section{Conclusion}

Within the impact parameter dependent saturated color dipole model, we have investigated the coherently production of very low-$p_{T}$ $J/\psi$ in peripheral nucleus-nucleus collisions at RHIC and LHC energies.  The scattering between the virtual photon and the nucleus is seen as the dissociation of virtual photon into a quark¨Cantiquark dipole with transverse size  followed by the interaction of color dipole with the proton via gluon exchanges. Our numerical results are consistent with experimental data for Au-Au and U-U collisions
at Relativistic Heavy Ion Collider(RHIC) energies.

%\textbf{Acknowledgements}

\section{Acknowledgements}

We wish to gratefully acknowledge many useful discussions with Larry D. McLerran. This work is supported by National Natural Science Foundation of China under Grant No. 12063006, 11805029, U1832105, 12005047, and Fundamental Research Funds for the Central Universities of China under Grant No. 3072020CFT1505.

\end{document}